\documentclass[12pt]{article}
\usepackage{amsmath,amssymb,amsfonts,maple2e,a4wide}

\DefineParaStyle{maplegroup}

\def\bar{\overline}
\def\Hl{\xi_\ell}
\def\Hn{\xi_n}
\def\Hm{\xi_m}
\def\Hmb{\xi_{\bar m}}
\def\mb{\bar m}

\begin{document}
\begin{center}
{\Large\bf Calculating symmetries in Newman-Tamburino metrics}
\end{center}

\medskip
{\bf John D.~Steele}

\smallskip
{\small School of Mathematics, University of New South Wales,
Sydney, New South Wales 2052, Australia.
 email: j.steele@unsw{.}edu{.}au}

\medskip
{\bf Abstract:} In this paper I 
show that the Newman-Tamburino spherical metrics 
always admit a Killing vector, correcting a claim by Collinson and French,
(1967  {\it J.~Math.~Phys.} {\bf 8} 701)
and also admit a homothety. A similar calculation is given for the limit of the Newman-Tamburino cylindrical metric.

\section{Introduction}

The Newman-Tamburino metrics are those vacuum solutions of the Einstein equations admitting hypersurface orthogonal geodesic rays with non-vanishing shear and divergence. In the Newman-Penrose formalism this implies that $\Psi_0=\kappa=0$, that $\rho$ is real and non-zero and $\sigma\neq0$.
In [1] Newman and Tamburino explicitly gave all such metrics and showed that they fall into two classes: the spherical, with $\rho^2\neq\sigma\overline\sigma$ and
the cylindical with  $\rho^2=\sigma\overline\sigma$. In [2] Collinson and French claimed to have shown that the former metrics admit at most one Killing vector, and that happens only in a particular subcase. 
In fact, the spherical Newman-Tamburino metrics always admit a Killing vector and also always admit
a homothety. This preprint is intended to show the full calculations and results when the homothetic equations of [3] are integrated for the Newman-Tamburino spherical metrics. The bulk of sections~\ref{sc:calcsph} and~\ref{sc:calclim} come from Maple 9 worksheets, exported to \TeX\ and suitably tidied up for better readability. 

Throughout I use the spin coefficient notation of [4]. For example
I use $\kappa'$, $\rho'$, $\sigma'$ and $\tau'$ in place of the more traditional $-\nu$, $-\mu$, $-\pi$ and $-\lambda$.

\section{Results}

The contravariant form of the Newman-Tamburino spherical metric [1] (see also [5], equation (26.21)) is
\begin{align*}
g^{22} &= -\frac{2r^2(\zeta\bar\zeta)^{1/2}}{R^2} + \frac{2rL}{A} +
\frac{2r^3A(\zeta^2+\bar\zeta^2)}{R^4} - 
\frac{4r^2A^2(\zeta\bar\zeta)^{3/2}}{R^4}
\\
g^{23} &= 4A^2(\zeta\bar\zeta)^{3/2}x\left[\frac{L}{2a^3} - 
\frac{r-2a}{2a^2R^2} -\frac{r-a}{R^4}\right]\\[2pt]
g^{24} &= 4A^2(\zeta\bar\zeta)^{3/2}y\left[\frac{L}{2a^3} - 
\frac{r+2a}{2a^2R^2} -\frac{r+a}{R^4}\right]\\
g^{33}&= -\frac{2(\zeta\bar\zeta)^{3/2}}{(r+a)^2}\qquad
g^{44}= -\frac{2(\zeta\bar\zeta)^{3/2}}{(r-a)^2}\qquad
g^{12}=1.
\end{align*} 
Here our coordinates are $x^1=u$, $x^2=r$, $x^3+ix^4=x+iy=\zeta$ and
$$ A(u) = bu+c,\qquad L=\frac{1}{2}\log\left(\frac{r+a}{r-a}\right)\qquad
a=A(\zeta\bar\zeta)^{1/2}\qquad R^2=r^2-a^2.$$
Here $b$ and $c$ are real constants.

The Collinson and French result (also quoted in [5]) is that there is a
Killing vector only in the case where $A$ is constant --- in this situation the Killing vector is the obvious $\partial_u$.
However, if $b\neq0$ we can set $c=0$ by a coordinate change and then the vector 
$$K^a=-u\partial_u+r\partial_r +2x\partial_x+2y\partial_y.$$
is a Killing vector, as will be shown in section~\ref{sc:calcsph}.
This can be checked directly: consider the flow of $K^a$. This scales the
coordinates by
$$ u\rightarrow \lambda^{-1}u,\qquad r\rightarrow \lambda r,\qquad 
\zeta\rightarrow \lambda^2\zeta$$
for real parameter $\lambda>0$. Under this scaling it is easy to check that all the contravariant components given above are homogeneous in $\lambda$ (when $A=bu$), and all of the correct degree to make the flow isometric. For example, the 
$g^{22}$ component is homogeneous of degree 2, and so the metric term
$g^{22}\dfrac{\partial}{\partial r}\otimes\dfrac{\partial}{\partial r}$ 
is unchanged under the flow.

Also the vector
$$H = r\partial_r +x\partial_x+y\partial_y$$
is a homothety, whatever $A$ is (see section~\ref{sc:calcsph}).
Alternatively, the flow of $H$ is
 $$ u\rightarrow u,\qquad r\rightarrow \lambda r,\qquad 
\zeta\rightarrow \lambda\zeta,$$
and we again find that all the contravariant components given above are homogeneous in $\lambda$, and all of the correct degree to make the flow homothetic. For example, the 
$g^{22}$ component is homogeneous of degree 1, and so the metric term
$g^{22}\dfrac{\partial}{\partial r}\otimes\dfrac{\partial}{\partial r}$ 
scales by $\lambda^{-1}$: the same scaling applies to all the metric terms.

\medskip
Newman and Tamburino [1] also give the following metric,
which arises as a limit of the cylindrical case
(see also [5] (26.23) for corrections to the $du^2$ coefficient):
$$ds^2 = 2\,du\,dr -x^{-2}\left[b+\log(r^2x^4)\right]du^2 +4\frac{r}{x}du\,dx
-r^2dx^2 - x^2dy^2,
$$
with the same coordinates as used in the spherical case.
The Killing vectors here are obvious ($\partial_u$ and $\partial_y$) and 
as we shall see there is also a homothetic vector (see section~\ref{sc:calclim})
$$H_2 = 2r\partial_r-x\partial_x+2y\partial_y.$$
One can use the flow of $H_2$ to check it is a homothety as well.

\section{The Calculations (spherical case)}\label{sc:calcsph}

The basic information is taken from Collinson and French [2], and Newman and Tamburino [1]. See those papers for those spin coefficients that are not actually calculated here. I have checked in a separate calculation that their results are correct as quoted. I use as coordinates $u$, $r$, $\zeta=x+iy$. 

Collinson and French [2] wrote the conformal Killing equations in 
Newman-Penrose form and used that in their work, although there are a few minor typos in their paper. Here, I will use the formalism of [3], which generalised the ideas of [6] into a form suitable for this task. I will use to the notation of [3] for the components of the homothety 
$$\xi_a =\Hn\ell_a + \Hl n_a - \Hmb m_a - \Hm \mb_a,$$ 
and its bivector, $F_{ab}$, with anti-self dual 
$$
{}^{-}{F}_{ab} =2\phi_{00}\,\ell_{[a}m_{b]}
 + 2\phi_{01}\,(\ell_{[a}n_{b]} - m_{[a}\mb_{b]})
 - 2\phi_{11}\,n_{[a}\mb_{b]}.
$$ 
The tetrad is a standard tetrad (see [1]), based around the 
Debever-Penrose vector $\ell^a=\partial_r$, see [1] and [2] for further detail. 
Since the tetrad is normalised, for the Penrose-Rindler spin coefficients used in [3]
we have $\gamma'=-\epsilon$, $\beta'=-\alpha$ etc. 

In the Maple I use use {\tt z} for $\zeta$ and {\tt w} for $\bar\zeta$;
{\tt Hl} for $\Hl$ etc. I typically add a {\tt b} for 
a complex conjugate ({\tt Hmb} is $\Hmb$) and a {\tt 1} for a dash ({\tt rho1}
is $\rho'$). 

 Firstly, define the terms {\tt a}, {\tt a0} (the latter is $\alpha_0$ in [2]).

\begin{mapleinput}
\mapleinline{active}{1d}{a:=A(u)*z^(1/2)*w^(1/2):}{}
\end{mapleinput}\begin{mapleinput}
\mapleinline{active}{1d}{a0:=3/4*w^(3/4)*z^(-1/4):}{}
\end{mapleinput}\begin{mapleinput}
\mapleinline{active}{1d}{a0b:=3/4*z^(3/4)*w^(-1/4):}{}
\end{mapleinput} 

Rather than use the explicit definitions for $L$ and $R$ in [2] and [1], I will leave them as ``unknown'' functions and define a routine later that will substitute for their derivatives. I will also use $Q(u,r,z,w)$ in place of 
$1/R^2$ to make things more transparent. I define what these functions actually are so we can substitute for them more easily when that become useful. I also define dummy symbols to use in place of the full functional dependence of
$L$ and $Q$ for ease of readability. I have also suppressed the functional
dependence in the Maple output, replacing $Q(u,r,z,w)$ with $Q(x^a)$ for example.
\begin{mapleinput}
\mapleinline{active}{1d}{LL:=L(u,r,z,w):Lis:=1/2*log((r+a)/(r-a));}{}
\end{mapleinput}
\[
{\it Lis}\, := \frac{1}{2}\ln  \left( {\frac {r+A \left( u \right) \sqrt {z}\sqrt {w}}{r-A \left( u \right) \sqrt {z}\sqrt {w}}} \right) 
\]
\begin{mapleinput}
\mapleinline{active}{1d}{QQ:=Q(u,r,z,w):Qis:=1/(r^2-a^2);}{}
\end{mapleinput}
\mapleresult
\begin{maplelatex}
\mapleinline{inert}{2d}{Qis := 1/(r^2-A(u)^2*z*w)}{%
\[
{\it Qis}\, := \, \left( {r}^{2}- \left( A \left( u \right)  \right) ^{2}zw \right) ^{-1}
\]
}

\end{maplelatex}
 Now we define the routine to simplify derivatives and products and also add a line to collect terms.
\begin{mapleinput}
\mapleinline{active}{1d}{diffsbs:=proc(XX)
subs(diff(L(u,r,z,w),r)=-a*Q(u,r,z,w),XX):
subs(diff(L(u,r,z,w),u)=r*Q(u,r,z,w)*diff(a,u),
subs(diff(L(u,r,z,w),w)=r*Q(u,r,z,w)*diff(a,w),
subs(diff(L(u,r,z,w),z)=r*Q(u,r,z,w)*diff(a,z),
subs(diff(Q(u,r,z,w),r)=-2*r*Q(u,r,z,w)^2,
subs(diff(Q(u,r,z,w),z)=2*Q(u,r,z,w)^2*a*diff(a,z),
subs(diff(Q(u,r,z,w),u)=2*Q(u,r,z,w)^2*a*diff(a,u),
subs(diff(Q(u,r,z,w),w)=2*Q(u,r,z,w)^2*a*diff(a,w),
student[powsubs](r^2=a^2+1/Q(u,r,z,w),expand(
collect(
end proc:}{}
\end{mapleinput} 
The terms {\tt S} and {\tt Sb} are  $\psi_0^1$ and its conjugate in [2].
\begin{mapleinput}
\mapleinline{active}{1d}{S:=2*A(u)^2*z^(3/4)*w^(3/4)*z;Sb:=2*A(u)^2*z^(3/4)*w^(3/4)*w:}{}
\end{mapleinput}
\mapleresult
\begin{maplelatex}
\mapleinline{inert}{2d}{S := 2*A(u)^2*z^(7/4)*w^(3/4)}{%
\[
S\, := \,2\, A \left( u \right)^{2}{z}^{7/4}{w}^{3/4}
\]
}

\end{maplelatex}
 And the curvature component $\Psi_1$ is given in [1]. \begin{mapleinput}
\mapleinline{active}{1d}{Psi1:=S*QQ^2;}{}
\end{mapleinput}
\mapleresult
\begin{maplelatex}
\mapleinline{inert}{2d}{Psi1 := 2*A(u)^2*z^(7/4)*w^(3/4)*Q(x^a)^2}{%
\[
{\it \Psi 1}\, := \,2\,A \left( u \right)^{2}{z}^{7/4}{w}^{3/4} Q\left(x^a\right)  ^{2}
\]
}

\end{maplelatex}
Now from [2] we have $\kappa=\epsilon=\tau'=\Psi_0=0$, and $\rho$ and $\sigma$ real --- these  can also be easily checked by Maple. So by [3]($6a$) $D\Hl=0$. Using $\tau=\bar\alpha+\beta$ ([2]), [3]($6c$) becomes 
$$ \delta\Hl=\tau\Hl-\rho\Hm-\sigma\Hmb-\phi_{11}.$$
We next use equation [3](11), since $\ell^a$ is a Debever-Penrose 
vector. Unfortunately, [3](11) contains an error --- the right hand side
is the complex conjugate of what it ought to be. With this correction, we have
$$ \phi_{11} = -\tau\Hl+\rho\Hm+\sigma\Hmb.$$
Hence $\delta\Hl=0$ and $\Hl=\Hl(u)$, as found in [2]. 

Equation ($10a$) of [3] is
$$ D\phi_{11} = -\Hl\Psi_1 ,
$$ and so integrates to give the $r$ dependence of $\phi_{11}$, 
here called {\tt p11}. We ignore the factor independent of $r$ when integrating:
\begin{mapleinput}
\mapleinline{active}{1d}{int(-Qis^2,r);}{}
\end{mapleinput}
\[
 -\,{\frac {r}{2 A \left( u \right)^{2}zw \left({r}^{2}-A(u)^{2}zw \right) }}+\frac{1}{2}{\rm arctanh} \left( {\frac {r}{A \left( u \right) \sqrt {zw}}} \right) A\left(u\right)^{-3} {z}^{-1}{w}^{-1}{\frac {1}{\sqrt {zw}}} 
\]
The arctanh term here is just $L$ and we get
\begin{mapleinput}
\mapleinline{active}{1d}{p11:=S*Hl(u)/2/a^2*r*QQ-S*Hl(u)/2/a^3*LL+p110(u,z,w);}{}
\end{mapleinput}
\[
{\it p11}\, := \,{\frac {{z}^{3/4}{\it Hl} \left( u \right) rQ \left(x^a\right) }{{w}^{1/4}}}-{\frac {{z}^{1/4}{\it Hl} \left( u \right) L \left(x^a\right) }{A \left( u \right) {w}^{3/4}}}+{\it p110}\left( u,z,w \right) 
\]
Here {\tt p110(u,z,w)} is the integration constant.
 Now the spin coefficients  --- see [2].
\begin{mapleinput}
\mapleinline{active}{1d}{alpha:=expand(simplify(S*LL*r/2/a^2*QQ+QQ*(r*a0+a*a0b)-S*QQ/a/2,radical)):}{}
\end{mapleinput}
\begin{mapleinput}
\mapleinline{active}{1d}{beta:=-S*LL*QQ/2/a-QQ*(r*a0b+a*a0):}{}
\end{mapleinput}   Now a routine to take conjugates nicely, as we need conjugates to define $\tau$.
\begin{mapleinput}
\mapleinline{active}{1d}{conj:=proc(XX)
 subs(z=w1,w=z1,XX):
subs(w1=w,z1=z,
subs(L(u,r,w,z)=L(u,r,z,w),Q(u,r,w,z)=Q(u,r,z,w),
subs(p110(u,w,z)=p110b(u,z,w),p110b(u,w,z)=p110(u,z,w),
end proc:}{}
\end{mapleinput}\begin{mapleinput}
\mapleinline{active}{1d}{conj(alpha)+beta:}{}
\end{mapleinput}\begin{mapleinput}
\mapleinline{active}{1d}{tau:=collect(
\end{mapleinput}
\mapleresult
\begin{maplelatex}
\mapleinline{inert}{2d}{tau := (-A(u)*z^(5/4)*w^(1/4)+r*w^(3/4)/z^(1/4))*Q(u, r, z, w)*L(u, r, z, w)-A(u)*w^(5/4)*z^(1/4)*Q(u, r, z, w)}{%
\[
\tau\, := \, \left( -A \left( u \right) {z}^{5/4}{w}^{1/4}+ {r{w}^{3/4}}{{z}^{-1/4}} \right) Q \left(x^a\right) L \left(x^a\right) -
A \left( u \right) {w}^{5/4}{z}^{1/4}Q \left(x^a \right) 
\]
}

\end{maplelatex}
\begin{mapleinput}
\mapleinline{active}{1d}{rho:=-r*Q(u,r,z,w):}{}
\end{mapleinput}\begin{mapleinput}
\mapleinline{active}{1d}{sigma:=a*Q(u,r,z,w):}{}
\end{mapleinput}\begin{mapleinput}
\mapleinline{active}{1d}{sigma1:=r*LL^2*S^2/4/a^4*QQ+r*LL*S*Sb/2/a^4*QQ-r*diff(a,u)*QQ-S*Sb/2/a^3*QQ:}{}
\end{mapleinput}\begin{mapleinput}
\mapleinline{active}{1d}{unprotect(gamma);}{}
\end{mapleinput}\begin{mapleinput}
\mapleinline{active}{1d}{gamma:=-r*LL^2*S*Sb/4/a^4*QQ+S^2*LL^2/4/a^3*QQ+a*r*LL*(S*a0b-Sb*a0)/2/a^3*QQ
 + S*Sb*QQ*LL/2/a^3 - S*Sb/4/a^3*(LL/2/a^2-r/2/a*QQ) +
(S^2/2/a-a*(S*a0b-Sb*a0))/2/a^2*QQ:}{}
\end{mapleinput}   

We need derivative operators $\delta$ and $\delta'$ to find $\rho'$. Firstly, the components of $m^a$ come from [2] and [1].
\begin{mapleinput}
\mapleinline{active}{1d}{om0:=-A(u)*z^(1/4)*w^(5/4):}{}
\end{mapleinput}\begin{mapleinput}
\mapleinline{active}{1d}{omega:=-Sb*LL/2/a^2+(r*om0-a*conj(om0))*QQ;}{}
\end{mapleinput}
\mapleresult
\begin{maplelatex}
\mapleinline{inert}{2d}{omega := -w^(3/4)*L(u, r, z, w)/z^(1/4)+(-r*A(u)*z^(1/4)*w^(5/4)+A(u)^2*z^(7/4)*w^(3/4))*Q(u, r, z, w)}{%
\[
\omega\, := \,-{w}^{3/4}{z}^{-1/4} L \left(x^a\right) + \left( -rA \left( u \right){z}^{1/4}{w}^{5/4}+  A \left( u \right)^{2}{z}^{7/4}{w}^{3/4} \right) Q \left(x^a\right) 
\]
}

\end{maplelatex}
\begin{mapleinput}
\mapleinline{active}{1d}{omega1:=conj(omega):}{}
\end{mapleinput}\begin{mapleinput}
\mapleinline{active}{1d}{P:=z^(3/4)*w^(3/4):}{}
\end{mapleinput}\begin{mapleinput}
\mapleinline{active}{1d}{del:=XX->omega*diff(XX,r)+2*r*P*QQ*diff(XX,w)-2*a*P*QQ*diff(XX,z):}{}
\end{mapleinput}\begin{mapleinput}
\mapleinline{active}{1d}{del1:=XX->omega1*diff(XX,r)+2*r*P*QQ*diff(XX,z)-2*a*P*QQ*diff(XX,w):}{}
\end{mapleinput} 
To calculate $\rho'$ we use [4] (4.11.12 $e'$).
\begin{mapleinput}
\mapleinline{active}{1d}{expand((diff(sigma1,r)-rho*sigma1)/sigma): }{}
\end{mapleinput}\begin{mapleinput}
\mapleinline{active}{1d}{rho1:=diffsbs(
\end{mapleinput}
\begin{align*}
{\it \rho 1}&:= \,-A \left( u \right) {z}^{2}L\left(x^a\right)^{2} Q\left(x^a\right) -2\left(A \left( u \right) zw+{\frac {{z}^{3/2}r}{\sqrt {w}}} \right) Q \left(x^a\right) L(x^a)+A \left( u \right) zw\, {\frac {d\,A}{du}}\, Q \left(x^a\right)
\end{align*}

We check some curvature equations next before we go on.
\begin{mapleinput}
\mapleinline{active}{1d}{diffsbs(diff(tau,r) - rho*tau-sigma*conj(tau)-Psi1); #[4]4.11.12c}{}
\end{mapleinput}
\mapleresult
\begin{maplelatex}
\mapleinline{inert}{2d}{0}{%
\[
0
\]
}

\end{maplelatex}
\begin{mapleinput}
\mapleinline{active}{1d}{diffsbs(diff(alpha,r)-rho*alpha-sigma*beta); # [4] 4.11.12h & i';}{}
\end{mapleinput}
\mapleresult
\begin{maplelatex}
\mapleinline{inert}{2d}{0}{%
\[
0
\]
}

\end{maplelatex}
\begin{mapleinput}
\mapleinline{active}{1d}{diffsbs(diff(beta,r)-sigma*alpha-beta*rho-Psi1);  #[4]4.11.12h'& i;}{}
\end{mapleinput}
\mapleresult
\begin{maplelatex}
\mapleinline{inert}{2d}{0}{%
\[
0
\]
}

\end{maplelatex}
\begin{mapleinput}
\mapleinline{active}{1d}{Psi2:=-diffsbs(diff(rho1,r)-rho1*rho-sigma*sgma1) ; #  [4] 4.11.12 f'}{}
\end{mapleinput}
\mapleresult
\begin{maplelatex}
\mapleinline{inert}{2d}{Psi2 := -4*A(u)^2*z^(5/2)*L(u, r, z, w)*Q(u, r, z, w)^2*w^(1/2)-(2*A(u)*z^2*r+4*A(u)^2*z^(3/2)*w^(3/2))*Q(u, r, z, w)^2}{%
\[
{\it \Psi 2}\, := \,-4 A \left( u \right)^{2}{z}^{5/2}L \left(x^a\right)  
Q\left(x^a\right)^{2}\sqrt {w}- \left( 2\,A \left( u \right) {z}^{2}r+4A \left( u \right)^{2}{z}^{3/2}{w}^{3/2}\right)Q \left(x^a\right) ^{2}
\]
}
\end{maplelatex}
This expression for $\Psi_2$ agrees with [1].
\begin{mapleinput}
\mapleinline{active}{1d}{diffsbs(diff(gamma,r) -beta*conj(tau)-alpha*tau-Psi2); # [4]4.11.12k}{}
\end{mapleinput}
\mapleresult
\begin{maplelatex}
\mapleinline{inert}{2d}{0}{%
\[
0
\]
}

\end{maplelatex}
\begin{mapleinput}
\mapleinline{active}{1d}{diffsbs(del(rho)-del1(sigma)-rho*(conj(alpha)+beta)+
sigma*(3*alpha-conj(beta) ) + Psi1); # [4] 4.11.12 d}{}
\end{mapleinput}
\mapleresult
\begin{maplelatex}
\mapleinline{inert}{2d}{0}{%
\[
0
\]
}

\end{maplelatex}
\begin{mapleinput}
\mapleinline{active}{1d}{diffsbs( del1(beta)-del(alpha)-rho*rho1+sigma*sigma1+alpha*conj(alpha)+
beta*conj(beta)-2*alpha*beta-Psi2  ); # [4] 4.11.12 l}{}
\end{mapleinput}
\mapleresult
\begin{maplelatex}
\mapleinline{inert}{2d}{0}{%
\[
0
\]
}

\end{maplelatex}
 Integrating [3] ($6g$) and using [3](11) (corrected, see above):
\begin{mapleinput}
\mapleinline{active}{1d}{Hm1:=-r*p110(u,z,w) + Hl(u)*S/2/a^3*r*LL+Hm0;}{}
\end{mapleinput}
\[
{\it Hm1}\, := \,-r{\it p110} \left( u,z,w \right) +{\frac {{\it Hl} \left( u \right) z^{1/4}rL \left(x^a\right) }{A \left( u \right) {w}^{3/4}}}+{\it Hm0}
\]

\begin{mapleinput}
\mapleinline{active}{1d}{Hmb1:=-r*p110b(u,z,w) + Hl(u)*Sb/2/a^3*r*LL+Hmb0;}{}
\end{mapleinput}
\[
{\it Hmb1}\, := \,-r{\it p110b} \left( u,z,w \right) +{\frac {{\it Hl} \left( u \right) {w}^{1/4}rL \left( x^a\right) }{A \left( u \right) {z}^{3/4}}}+{\it Hmb0}
\]

 By [3] (11) the following ought to be zero.
\begin{mapleinput}
\mapleinline{active}{1d}{diffsbs(p11+tau*Hl(u)-rho*Hm1-sigma*Hmb1):}{}
\end{mapleinput}\begin{mapleinput}
\mapleinline{active}{1d}{collect(
\end{mapleinput}
\begin{align*}
 &\left( {\it Hm0}+{\frac {{z}^{3/4}{\it Hl} \left( u \right) }{{w}^{1/4}}}+A \left( u \right) \sqrt {z}\sqrt {w}\,{\it p110b} \left( u,z,w \right)  \right) r\\
& -A \left( u \right) \sqrt {z}\sqrt {w}{\it Hmb0}- \left( A \left( u \right)  \right) ^{2}zw\,{\it p110} \left( u,z,w \right) -{\it Hl} \left( u \right) A \left( u \right) {w}^{5/4}{z}^{1/4}
\end{align*}
\begin{mapleinput}
\mapleinline{active}{1d}{expand(solve(coeff(
\end{mapleinput}
\[
-{\frac {{z}^{3/4}{\it Hl} \left( u \right) }{{w}^{1/4}}}-A \left( u \right) \sqrt {z}\sqrt {w}{\it p110b} \left( u,z,w \right) ,\,-\sqrt {z}\sqrt {w}A \left( u \right) {\it p110} \left( u,z,w \right) -{\frac {{w}^{3/4}{\it Hl} \left( u \right) }{{z}^{1/4}}}
\]
 So we get
\begin{mapleinput}
\mapleinline{active}{1d}{Hm:=-r*p110(u,z,w)-a*p110b(u,z,w)+Hl(u)*expand(S/2/a^3*(r*LL-a)):}{}
\end{mapleinput}\begin{mapleinput}
\mapleinline{active}{1d}{Hmb:=conj(Hm):}{}
\end{mapleinput}  
These agree with the components in [2] (their $V_3$ and $V_4$). 
Now we use [3] (10b) to get $\phi_{01}$.
\begin{mapleinput}
\mapleinline{active}{1d}{diffsbs(Psi1*Hm/2/sigma-beta*p11/sigma+del(p11)/2/sigma):}{}
\end{mapleinput}
\begin{mapleinput}
\mapleinline{active}{1d}{p01:=collect(
\end{mapleinput}
\begin{align*}
{\it p01} &:= \left(\left( \frac{1}{2}\,\sqrt {z}\sqrt {w}r+ \left( 2\,{\frac {{z}^{3/2}r}{\sqrt {w}}}+A \left( u \right) zw \right) L \left(x^a\right)  \right) Q \left( x^a\right) -{\frac {z \left( L \left(x^a\right)  \right) ^{2}}{A \left( u \right) w}}-{\frac {L \left(x^a\right) }{2 A \left( u \right) }} \right) {\it Hl} \left( u \right)\\
&\quad +{\frac {{z}^{3/4}L \left(x^a\right) {\it p110} \left( u,z,w \right) }{{w}^{1/4}}}+
{\frac {3{w}^{3/4}{\it p110}\left( u,z,w \right)}{4{z}^{1/4}}}\\ &  {} -
\Bigl(A \left( u \right) {z}^{5/4}{w}^{1/4}r\,{\it p110} \left( u,z,w \right)
+A \left( u \right)^{2}{z}^{7/4}{w}^{3/4}{\it p110b} \left( u,z,w \right)\Bigr) 
Q\left(x^a \right)  \\
{} +& \left( {\frac {3{\it p110} \left( u,z,w \right) }{4A \left( u \right) {w}^{3/4}}} +
{\frac {{w}^{1/4}{\frac {\partial }{\partial w}}{\it p110} \left( u,z,w \right) }{A \left( u \right) }} \right) {z}^{1/4}r\mbox{}-{z}^{3/4}{w}^{3/4}{\frac {\partial }{\partial z}}{\it p110} \left( u,z,w \right) 
\end{align*}

 Now [3]($10 c$) and ($8a$) will give us information on the $w$ (that is, $\bar \zeta$) dependence of $\phi_{11}^0$.
\begin{mapleinput}
\mapleinline{active}{1d}{diffsbs(Psi1*Hmb-Psi2*Hl(u)+2*rho*p01+2*alpha*p11-del1(p11)); # [3]10c}{}
\end{mapleinput}
\[
 \left({\frac {3{\it p110} \left( u,z,w \right) }{2A \left( u \right) {w}^{3/4}}}-2\,{\frac {{w}^{1/4} }{A \left( u \right) }} {\frac {\partial }{\partial w}}{\it p110} \left( u,z,w \right)\right){z}^{1/4}
\]
\begin{mapleinput}
\mapleinline{active}{1d}{dsolve(
\end{mapleinput}
\[
{\it p110} \left( u,z,w \right) ={\frac {{\it \_F1} \left( u,z \right) }{{w}^{3/4}}}
\]
\begin{mapleinput}
\mapleinline{active}{1d}{diffsbs(diff(p01,r)+del1(p11)-2*rho*p01-2*alpha*p11); # [3] (8a)}{}
\end{mapleinput}
\[
 \left( {\frac {9{\it p110} \left( u,z,w \right) }{4A \left( u \right) {w}^{3/4}}}+3\,{\frac {{w}^{1/4}}{A \left( u \right) }} {\frac {\partial }{\partial w}}{\it p110} \left( u,z,w \right) \right) {z}^{1/4}
\]
\begin{mapleinput}
\mapleinline{active}{1d}{dsolve(
\end{mapleinput}
\[
{\it p110} \left( u,z,w \right) ={\frac {{\it \_F1} \left( u,z \right) }{{w}^{3/4}}}
\]
 Both giving the same result. Now we turn to $\Hn$ and [3]($6i$), which we solve for $\sigma \Hl$.
\begin{mapleinput}
\mapleinline{active}{1d}{rhs6i:=diffsbs((-del(Hm)-conj(sigma1)*Hl(u)-Hm*(conj(alpha)-beta))):}{}
\end{mapleinput}  
The imaginary part ought to be zero as $\Hn$ is real, so using results from [3] ($10c$) and ($8a$), we find the imaginary part divide out a common nor-zero factor and call what's left $X$.
\begin{mapleinput}
\mapleinline{active}{1d}{Imrhs6i:=
\end{mapleinput}\begin{mapleinput}
\mapleinline{active}{1d}{subs(p110(u,z,w)=F(u,z)/w^(3/4),p110b(u,z,w)=Fb(u,w)/z^(3/4),
\end{mapleinput}\begin{mapleinput}
\mapleinline{active}{1d}{X:=expand(
\end{mapleinput}
\[
X\, := \,-4\,{z}^{3/4}A \left( u \right) {\frac {\partial }{\partial z}}F \left( u,z \right) +4\,{w}^{3/4}A \left( u \right) {\frac {\partial }{\partial w}}{\it Fb} \left( u,w \right) +3\,{\frac {A \left( u \right) {\it Fb} \left( u,w \right) }{{w}^{1/4}}}-3\,{\frac {A \left( u \right) F \left( u,z \right) }{{z}^{1/4}}}
\]
Assuming $F$ is differentiable in $z$ we can split this
\begin{mapleinput}
\mapleinline{active}{1d}{subs(Fb=0,X):
\end{mapleinput}
\[
-4\,{z}^{3/4}A \left( u \right) {\frac {\partial }{\partial z}}F \left( u,z \right) -3\,{\frac {A \left( u \right) F \left( u,z \right) }{{z}^{1/4}}}
\]

 This is a (real) function of $u$ and $w$. We choose the shape of the separation function to simplify the solution to the differential equation slightly.
\begin{mapleinput}
\mapleinline{active}{1d}{dsolve(
\end{mapleinput}
\mapleresult
\begin{maplelatex}
\mapleinline{inert}{2d}{F(u, z) = (G(u)*z+_F1(u))/z^(3/4)}{%
\[
F \left( u,z \right) ={\frac {G \left( u \right) z+{\it \_F1} \left( u \right) }{{z}^{3/4}}}
\]
}

\end{maplelatex}
 Check this out:
\begin{mapleinput}
\mapleinline{active}{1d}{subs(F(u,z)=G(u)*z^(1/4)+H(u)/z^(3/4),Fb(u,w)=G(u)*w^(1/4)+Hb(u)/w^(3/4),X):
expand(
\end{mapleinput}
\mapleresult
\begin{maplelatex}
\mapleinline{inert}{2d}{0}{%
\[
0
\]
}

\end{maplelatex}
 So we define a simplification routine for $\phi_{11}^0$.
\begin{mapleinput}
\mapleinline{active}{1d}{P110sbs1:=proc(XX)
subs(p110(u,z,w)=F(u,z)/w^(3/4),p110b(u,z,w)=Fb(u,w)/z^(3/4),XX);
 subs(F(u,z)=G(u)*z^(1/4)+H(u)/z^(3/4),Fb(u,w)=G(u)*w^(1/4)+Hb(u)/w^(3/4),
expand(
end proc:}{}
\end{mapleinput} And check it works\begin{mapleinput}
\mapleinline{active}{1d}{P110sbs1(Imrhs6i);}{}
\end{mapleinput}
\mapleresult
\begin{maplelatex}
\mapleinline{inert}{2d}{0}{%
\[
0
\]
}

\end{maplelatex}
  Turning to [3]($6b$) next,
\begin{mapleinput}
\mapleinline{active}{1d}{eqn6b:=Hl(u)*(gamma+conj(gamma))-conj(tau)*Hm-tau*conj(Hm)-p01-conj(p01)+psi:}{}
\end{mapleinput}\begin{mapleinput}
\mapleinline{active}{1d}{P110sbs1(diffsbs(
}{}
\end{mapleinput}
\[
\psi-G \left( u \right) -3{\frac {H \left( u \right) }{2z}}-3{\frac {{\it Hb} \left( u \right) }{2w}}
\]
 This ought to be $\dot\Hl$, a function of $u$ only, so $H=0$ and we define a new simplification routine and test it out:
\begin{mapleinput}
\mapleinline{active}{1d}{P110sbs2:=proc(XX);
 expand( subs( p110(u,z,w)= (psi-diff(Hl(u),u))*z^(1/4)/w^(3/4) ,
 p110b(u,z,w)=(psi-diff(Hl(u),u))*w^(1/4)/z^(3/4),XX));
collect(
}{}
\end{mapleinput}\begin{mapleinput}
\mapleinline{active}{1d}{P110sbs2(diffsbs(eqn6b));}{}
\end{mapleinput}
\mapleresult
\begin{maplelatex}
\mapleinline{inert}{2d}{diff(Hl(u), u)}{%
\[
{\frac {d}{du}}{\it Hl} \left( u \right) 
\]
}

\end{maplelatex}
This is as it should be. Now we can  define $\Hn$.
\begin{mapleinput}
\mapleinline{active}{1d}{P110sbs2(diffsbs(rhs6i+conj(rhs6i))/sigma/2):}{}
\end{mapleinput}\begin{mapleinput}
\mapleinline{active}{1d}{Hn:=collect(
\end{mapleinput} 
We check this against the [2] version, called $V_2$ there. 
It is clear from the shape of $\Hm$ ($=V_3$ of [2]) that $a_0$ in [2] is my $\phi_{11}^0$.
\begin{mapleinput}
\mapleinline{active}{1d}{ay0:=p110(u,z,w):ay0b:=p110b(u,z,w):}{}
\end{mapleinput}\begin{mapleinput}
\mapleinline{active}{1d}{V2:=r*LL^2*(-S^2-Sb^2)*Hl(u)/4/a^5 - LL^2*S*Sb*Hl(u)/4/a^4 - 
LL*(ay0b*S+ay0*Sb)/2/a + r*(2*a*Hl(u)*diff(a,u)-ay0*Sb-ay0b*S)/2/a^2 +
 (-2*a^3*ay0*S-2*a^3*ay0b*Sb+Hl(u)*S*Sb)/4/a^4+r*LL*(-2*a^3*ay0*S-
2*a^3*ay0b*Sb- Hl(u)*S*Sb)/4/a^5 +  1/QQ*(-3*ay0*S-3*ay0b*Sb-
4*a^2*2*P*(diff(p110(u,z,w),w) +diff(p110b(u,z,w),z) ) )/8/a^3:}{}
\end{mapleinput}\begin{mapleinput}
\mapleinline{active}{1d}{expand(Hn-P110sbs2(V2)):}{}
\end{mapleinput}\begin{mapleinput}
\mapleinline{active}{1d}{simplify(subs(psi=0,diff(Hl(u),u)=0,

}{}
\end{mapleinput}
\mapleresult
\begin{maplelatex}
\mapleinline{inert}{2d}{0}{%
\[
0
\]
}

\end{maplelatex}
  So our $\Hn$ agrees with [2] in the case of their Killing vector ($\psi=0$ 
and $\Hl$ constant). However, if $\Hl$ is not constant, the terms differ:
\begin{mapleinput}
\mapleinline{active}{1d}{simplify(subs(psi=0,
\end{mapleinput}
$$-\frac{2}{{z}^{3/2}{w}^{3/2}}\,\frac {d{\it Hl}\left(u\right)}{du}
 \left( 
\left( {w}^{3}z+ {z}^{3}w+2 {z}^{2}{w}^{2}L \left( x^a \right) \right)A \left( u \right)+
L \left(x^a\right) r({z}^{5/2}\sqrt{w}+{w}^{5/2}\sqrt{z}) +2r{z}^{3/2}{w}^{3/2}
 \right) $$

Next, we put our $\Hn$ into [3]($6d$).
\begin{mapleinput}
\mapleinline{active}{1d}{eqn6d:=diffsbs(diff(Hn,r)-p01-conj(p01)-psi):}{}
\end{mapleinput}\begin{mapleinput}
\mapleinline{active}{1d}{P110sbs2(
\end{mapleinput}
\[
{\frac {{\it Hl} \left( u \right) {\frac {d}{du}}A \left( u \right) }{A \left( u \right) }}-{\frac {d}{du}}{\it Hl} \left( u \right) 
\]
\begin{mapleinput}
\mapleinline{active}{1d}{dsolve(
\end{mapleinput}
\mapleresult
\begin{maplelatex}
\mapleinline{inert}{2d}{Hl(u) = _C1*A(u)}{%
\[
{\it Hl} \left( u \right) ={\it \_C1}\,A \left( u \right) 
\]
}

\end{maplelatex}
  So next a routine to replace $\Hl(u)$ with a multiple of $A(u)$, 
and also to kill off the second derivative of $A(u)$.
\begin{mapleinput}
\mapleinline{active}{1d}{ Hlsbs:=proc(XX)
 subs(Hl(u)=C*A(u),XX);subs(diff(A(u),u,u)=0,
end proc:}{}
\end{mapleinput}

 We now try [3]($6j$).
\begin{mapleinput}
\mapleinline{active}{1d}{del1(Hm)+conj(rho1)*Hl(u)+rho*Hn+(conj(beta)-alpha)*Hm-p01+conj(p01)+psi: }{}
\end{mapleinput}\begin{mapleinput}
\mapleinline{active}{1d}{Hlsbs(P110sbs2(diffsbs(
\end{mapleinput}
\mapleresult
\begin{maplelatex}
\mapleinline{inert}{2d}{0}{%
\[
0
\]
}

\end{maplelatex}
 So that is satisfied. Now for $\phi_{00}$, 
which we get from the conjugate of [3]($6f$).
\begin{mapleinput}
\mapleinline{active}{1d}{eqn6f:=-del1(Hn)-(conj(beta)+alpha)*Hn-conj(rho1)*Hmb-sigma1*Hm:}{}
\end{mapleinput}\begin{mapleinput}
\mapleinline{active}{1d}{p00:=diffsbs(P110sbs2(diffsbs(
\end{mapleinput} 
I've supressed this component as it's very long, but we check the result with [3] ($8d$).\begin{mapleinput}
\mapleinline{active}{1d}{del1(p01)+diff(p00,r)-rho*p00-sigma1*p11:}{}
\end{mapleinput}\begin{mapleinput}
\mapleinline{active}{1d}{P110sbs2(diffsbs(
\end{mapleinput}
\mapleresult
\begin{maplelatex}
\mapleinline{inert}{2d}{0}{%
\[
0
\]
}

\end{maplelatex}
 To go any further we need to get the components of 
$n^a=(1,U,X^3,X^4)$ and to define $D'$. 
Taking the metric terms from [1] and [2]:\begin{mapleinput}
\mapleinline{active}{1d}{gup22:=-2*r^2*sqrt(w)*sqrt(z)*QQ + 2*r*LL/A(u) + QQ^2*(2*r^3*A(u)*(w^2+z^2) - 4*r^2*A(u)^2*w^(3/2)*z^(3/2)):}{}
\end{mapleinput}\begin{mapleinput}
\mapleinline{active}{1d}{gup22+2*omega*conj(omega):}{}
\end{mapleinput}\begin{mapleinput}
\mapleinline{active}{1d}{U:=diffsbs(
\end{mapleinput}
\begin{align*}
U:&= \,\sqrt {z}\sqrt {w}L \left(x^a\right)^{2}+
\left( \left({w}^{2}+{z}^{2}\right)A \left( u \right)  r-2\, A \left( u \right)^{2}{z}^{3/2}{w}^{3/2} \right) Q \left(x^a\right) -\sqrt {z}\sqrt {w}
 \\ & + \left( \left( 
- A \left( u \right)^{2}{z}^{5/2}\sqrt {w}+2\,zwrA \left( u \right) - A\left( u \right)^{2}\sqrt {z}{w}^{5/2} \right) Q \left(x^a \right) +{\frac {r}{A \left( u \right) }} \right) L \left(x^a\right) 
\end{align*}
\begin{mapleinput}
\mapleinline{active}{1d}{gup33:=-2*z^(3/2)*w^(3/2)/(r+a)^2:gup44:=-2*z^(3/2)*w^(3/2)/(r-a)^2:}{}
\end{mapleinput}  These next two terms are the components of $m^a$.
\begin{mapleinput}
\mapleinline{active}{1d}{xi3:=P*(r-a)*QQ;xi4:=I*P*(r+a)*QQ;}{}
\end{mapleinput}
\mapleresult
\begin{maplelatex}
\mapleinline{inert}{2d}{xi3 := z^(3/4)*w^(3/4)*(r-A(u)*z^(1/2)*w^(1/2))*Q(x^a)}{%
\[
{\it xi3}\, := \,{z}^{3/4}{w}^{3/4} \left( r-A \left( u \right) \sqrt {z}\sqrt {w} \right) Q \left(x^a\right) 
\]
}
\mapleinline{inert}{2d}{xi4 := I*z^(3/4)*w^(3/4)*(r+A(u)*z^(1/2)*w^(1/2))*QQ}{%
\[
{\it xi4}\, := \,i{z}^{3/4}{w}^{3/4} \left( r+A \left( u \right) \sqrt {z}\sqrt {w} \right) Q \left(x^a \right) 
\]
}

\end{maplelatex}
\begin{mapleinput}
\mapleinline{active}{1d}{xi3*conj(xi4)+xi4*conj(xi3); # checking}{}
\end{mapleinput}
\mapleresult
\begin{maplelatex}
\mapleinline{inert}{2d}{0}{%
\[
0
\]
}

\end{maplelatex}
\begin{mapleinput}
\mapleinline{active}{1d}{simplify(( subs( Q(u,r,z,w)=Qis,xi3*conj(xi3)*2+gup33) ));}{}
\end{mapleinput}
\mapleresult
\begin{maplelatex}
\mapleinline{inert}{2d}{0}{%
\[
0
\]
}

\end{maplelatex}
\begin{mapleinput}
\mapleinline{active}{1d}{simplify(subs(Q(u,r,z,w)=Qis,xi4*conj(xi4)*2+gup44));}{}
\end{mapleinput}
\mapleresult
\begin{maplelatex}
\mapleinline{inert}{2d}{0}{%
\[
0
\]
}

\end{maplelatex}
\begin{mapleinput}
\mapleinline{active}{1d}{gup23:=4*A(u)^2*z^(3/2)*w^(3/2)*(z+w)/2*(LL/2/a^3-(r-2*a)*QQ/2/a^2
-(r-a)*QQ^2):}{}
\end{mapleinput}\begin{mapleinput}
\mapleinline{active}{1d}{gup24:=4*A(u)^2*z^(3/2)*w^(3/2)*(z-w)/2/I*(LL/2/a^3-(r+2*a)*QQ/2/a^2
-(r+a)*QQ^2):}{}
\end{mapleinput}\begin{mapleinput}
\mapleinline{active}{1d}{gup23+omega*conj(xi3)+conj(omega)*xi3:}{}
\end{mapleinput}\begin{mapleinput}
\mapleinline{active}{1d}{X3:=diffsbs(
\end{mapleinput}
\begin{align*}
{\it X3}&:= \, \left(  \left(  \left( -{z}^{3/2}\sqrt {w}-\sqrt {z}{w}^{3/2} \right) r+A \left( u \right) z{w}^{2}+A \left( u \right) {z}^{2}w \right) Q \left( x^a\right) +{\frac {w}{A \left( u \right) }}+{\frac {z}{A \left( u \right) }} \right) L \left(x^a\right) \\ 
&+ \left(  \left( -{z}^{3/2}\sqrt {w}-\sqrt {z}{w}^{3/2} \right) r+A \left( u \right) z{w}^{2}+A \left( u \right) {z}^{2}w \right) Q \left(x^a \right) 
\end{align*}
\begin{mapleinput}
\mapleinline{active}{1d}{gup24+omega*conj(xi4)+omega1*xi4:}{}
\end{mapleinput}\begin{mapleinput}
\mapleinline{active}{1d}{X4:=diffsbs(factor(
}{}
\end{mapleinput}
\begin{align*}
{\it X4} &:= i\left(  \left(  \left( -{z}^{3/2}\sqrt {w}+\sqrt {z}{w}^{3/2} \right) r+z{w}^{2}A \left( u \right) -{z}^{2}wA \left( u \right)  \right) Q \left( x^a \right) +{\frac {w}{A \left( u \right) }}-{\frac {z}{A \left( u \right) }} \right) L \left(x^a\right) \\
& + \left(  \left( -i\sqrt {z}{w}^{3/2}+i{z}^{3/2}\sqrt {w} \right) r+i{z}^{2}wA \left( u \right) -iz{w}^{2}A \left( u \right)  \right) Q \left(x^a \right) 
\end{align*}
 As a double check we firstly define the (contravariant) tetrad and then check against the metric terms.
\begin{mapleinput}
\mapleinline{active}{1d}{ell:=<0,1,0,0>:en:=<1,U,X3,X4>:}{}
\end{mapleinput}\begin{mapleinput}
\mapleinline{active}{1d}{em:=<0,omega,xi3,xi4>:emb:=map(conj,em):}{}
\end{mapleinput}\begin{mapleinput}
\mapleinline{active}{1d}{ell.Transpose(en)-em.Transpose(emb):}{}
\end{mapleinput}\begin{mapleinput}
\mapleinline{active}{1d}{
\end{mapleinput}\begin{mapleinput}
\mapleinline{active}{1d}{g:=map(diffsbs,
\end{mapleinput}\begin{mapleinput}
\mapleinline{active}{1d}{diffsbs(g[2,2]-gup22);}{}
\end{mapleinput}
\mapleresult
\begin{maplelatex}
\mapleinline{inert}{2d}{0}{%
\[
0
\]
}

\end{maplelatex}
\begin{mapleinput}
\mapleinline{active}{1d}{diffsbs(g[2,3]-gup23);}{}
\end{mapleinput}
\mapleresult
\begin{maplelatex}
\mapleinline{inert}{2d}{0}{%
\[
0
\]
}

\end{maplelatex}
\begin{mapleinput}
\mapleinline{active}{1d}{diffsbs(g[2,4]-gup24);}{}
\end{mapleinput}
\mapleresult
\begin{maplelatex}
\mapleinline{inert}{2d}{0}{%
\[
0
\]
}

\end{maplelatex}
\begin{mapleinput}
\mapleinline{active}{1d}{diffsbs(simplify(g[3,3]-gup33));}{}
\end{mapleinput}
\mapleresult
\begin{maplelatex}
\mapleinline{inert}{2d}{0}{%
\[
0
\]
}

\end{maplelatex}
\begin{mapleinput}
\mapleinline{active}{1d}{diffsbs(simplify(g[4,4]-gup44));}{}
\end{mapleinput}
\mapleresult
\begin{maplelatex}
\mapleinline{inert}{2d}{0}{%
\[
0
\]
}

\end{maplelatex}
 For a second check we apply  two of the commutators [4] (4.11.11) to $r$ and check what we get.\begin{mapleinput}
\mapleinline{active}{1d}{diff(U,r)+gamma+conj(gamma)-tau*conj(omega)-conj(tau)*omega: }{}
\end{mapleinput}\begin{mapleinput}
\mapleinline{active}{1d}{diffsbs(
\end{mapleinput}
\mapleresult
\begin{maplelatex}
\mapleinline{inert}{2d}{0}{%
\[
0
\]
}

\end{maplelatex}
\begin{mapleinput}
\mapleinline{active}{1d}{diff(X3,r)-tau*conj(xi3)-conj(tau)*xi3: }{}
\end{mapleinput}\begin{mapleinput}
\mapleinline{active}{1d}{diffsbs(
\end{mapleinput}
\mapleresult
\begin{maplelatex}
\mapleinline{inert}{2d}{0}{%
\[
0
\]
}

\end{maplelatex}
  Since all this checks out we go ahead and define $D'$.
\begin{mapleinput}
\mapleinline{active}{1d}{D1:=proc(XX)
 diff(XX,u)+diff(XX,r)*U+(X3+I*X4)*diff(XX,z)+(X3-I*X4)*diff(XX,w);
P110sbs2(diffsbs(
end proc:}{}
\end{mapleinput}  We make use of $D'$ firstly to find the last spin coefficient, $\kappa'$, using [4](4.11.12$g$).
\begin{mapleinput}
\mapleinline{active}{1d}{D1(beta)-del(gamma)-tau*rho1-alpha*conj(sigma1)-beta*(rho1+gamma-conj(gamma))
+gamma*(tau-beta-conj(alpha)): # should be -kappa'*sigma}{}
\end{mapleinput}\begin{mapleinput}
\mapleinline{active}{1d}{diffsbs(
\end{mapleinput}\begin{mapleinput}
\mapleinline{active}{1d}{kappa1:=-diffsbs(P110sbs2(
\end{mapleinput}
\begin{align*}
{\kappa 1}& := \,- \left( {z}^{5/4}{w}^{1/4}r-{\frac {{z}^{11/4}A \left( u \right) }{{w}^{1/4}}} \right)Q\left(x^a\right) L\left(x^a\right)^{3}\\
& - \left(\left( -2\,{\frac {{z}^{9/4}}{{w}^{3/4}}}+2\,{z}^{1/4}{w}^{5/4} \right) r-3\,{z}^{7/4}{w}^{3/4}A \left( u \right)  \right) Q \left(x^a \right)L \left( x^a\right)^{2} +\\
& \left(  \left({\frac {d\,A}{du}} {z}^{1/4}{w}^{5/4}-2\,{z}^{5/4}{w}^{1/4} \right) r-
A \left( u \right)\left({z}^{7/4}{w}^{3/4}{\frac {d\,A}{du}}
+4{z}^{3/4}{w}^{7/4}\right) \right) Q \left(x^a\right) L \left( x^a\right)\\
 & - \left( -2\,{z}^{7/4}{w}^{3/4}A \left( u \right) +{w}^{7/4}{z}^{3/4}A \left( u \right) {\frac {d\,A}{du}} -2\,{z}^{5/4}{w}^{1/4}r
{\frac {d\,A}{du}} \right) Q \left(x^a \right) 
\end{align*}
 Now to look at the equations that involve $D'$.  Firstly [3] ($6h$):
\begin{mapleinput}
\mapleinline{active}{1d}{eqn6h:=D1(Hm)+conj(kappa1)*Hl(u)+tau*Hn+(conj(gamma)-gamma)*Hm-conj(p00):}{}
\end{mapleinput}\begin{mapleinput}
\mapleinline{active}{1d}{P110sbs2(diffsbs(
\end{mapleinput}
\mapleresult
\begin{maplelatex}
\mapleinline{inert}{2d}{0}{%
\[
0
\]
}

\end{maplelatex}
 Then we look at [3] ($6e$):\begin{mapleinput}
\mapleinline{active}{1d}{eqn6e:=D1(Hn)+(gamma+conj(gamma))*Hn+kappa1*Hm+conj(kappa1)*Hmb:}{}
\end{mapleinput}\begin{mapleinput}
\mapleinline{active}{1d}{P110sbs2(diffsbs(
\end{mapleinput}\begin{mapleinput}
\mapleinline{active}{1d}{factor(Hlsbs(
\end{mapleinput}
\mapleresult
\begin{maplelatex}
\mapleinline{inert}{2d}{0}{%
\[
0
\]
}

\end{maplelatex}
and [3] ($8c$)\begin{mapleinput}
\mapleinline{active}{1d}{eq8c:=diffsbs(del(p01)+D1(p11)-sigma*p00-2*tau*p01-(rho1+2*gamma)*p11):}{}
\end{mapleinput}\begin{mapleinput}
\mapleinline{active}{1d}{factor(Hlsbs(P110sbs2(
}{}
\end{mapleinput}
\mapleresult
\begin{maplelatex}
\mapleinline{inert}{2d}{0}{%
\[
0
\]
}

\end{maplelatex}
and [3] ($8b$)\begin{mapleinput}
\mapleinline{active}{1d}{D1(p01)+del(p00)-(tau-2*beta)*p00-2*rho1*p01-kappa1*p11:}{}
\end{mapleinput}\begin{mapleinput}
\mapleinline{active}{1d}{P110sbs2(diffsbs(
\end{mapleinput}\begin{mapleinput}
\mapleinline{active}{1d}{factor(Hlsbs(
\end{mapleinput}
\mapleresult
\begin{maplelatex}
\mapleinline{inert}{2d}{0}{%
\[
0
\]
}

\end{maplelatex}
and finally [3] ($10d$).\begin{mapleinput}
\mapleinline{active}{1d}{eqn10d:=Psi2*Hm-Psi1*Hn-2*tau*p01-2*gamma*p11+D1(p11):}{}
\end{mapleinput}\begin{mapleinput}
\mapleinline{active}{1d}{P110sbs2(diffsbs(
\end{mapleinput}
\mapleresult
\begin{maplelatex}
\mapleinline{inert}{2d}{0}{%
\[
0
\]
}

\end{maplelatex}
Next we consider what happens if we have a Killing vector ($\psi=0$) with $\Hl$ zero ($C=0$)
\begin{mapleinput}
\mapleinline{active}{1d}{subs(C=0,psi=0,Hlsbs(P110sbs2(Hm)));
}{}
\end{mapleinput}
\mapleresult
\begin{maplelatex}
\mapleinline{inert}{2d}{0}{%
\[
0
\]
}

\end{maplelatex}
\begin{mapleinput}
\mapleinline{active}{1d}{subs(C=0,psi=0,Hlsbs(P110sbs2(Hn)));}{}
\end{mapleinput}
\mapleresult
\begin{maplelatex}
\mapleinline{inert}{2d}{0}{%
\[
0
\]
}

\end{maplelatex}
 Hence we cannot have both $\psi$ and $C$ zero. This is the Collinson and French result: only one Killing vector at most. We have a look at the homothety.
\begin{mapleinput}
\mapleinline{active}{1d}{Hl(u)*en+Hn*ell-Hm*emb-conj(Hm)*em:}{}
\end{mapleinput}\begin{mapleinput}
\mapleinline{active}{1d}{map(diffsbs,
\end{mapleinput}\begin{mapleinput}
\mapleinline{active}{1d}{map(P110sbs2,
\end{mapleinput}\begin{mapleinput}
\mapleinline{active}{1d}{map(Hlsbs,
\end{mapleinput}\begin{mapleinput}
\mapleinline{active}{1d}{subs(z=x+I*y,w=x-I*y,K):K:=map(expand,
\end{mapleinput}

\[
K := \begin{pmatrix}
 CA \left( u \right) \\ 
 -rC{\frac {d\,A}{du}}+2\,r\psi\\ 
 2\,\psi\,x-2\,C{\frac {d\,A}{du}} x\\ 
2\,\psi\,y-2\,C {\frac {d\,A}{du}} y
\end{pmatrix}
\]

So the obvious Killing vector if $A$ is constant:
\begin{mapleinput}
\mapleinline{active}{1d}{KK:=subs(psi=0,C=1/B,diff(A(u),u)=0,A(u)=B,K):
\end{mapleinput}
$$ \begin{pmatrix}
1 \\ 
0 \\ 
0 \\ 
0
\end{pmatrix} $$
The new Killing vector in the other case:
\begin{mapleinput}
\mapleinline{active}{1d}{KK2:=subs(psi=0,A(u)=u*B,C=1/B,K):map(simplify,
\end{mapleinput}
$$\begin{pmatrix}
 u\\ 
 -r\\ 
 -2x\\ 
-2y
\end{pmatrix} 
$$

And the proper homothety for both cases:
\begin{mapleinput}
\mapleinline{active}{1d}{HH:=subs(C=0,psi=1,K);}{}
\end{mapleinput}
$$\begin{pmatrix}
 0\\ 
 2r\\ 
 2x\\ 
2y
\end{pmatrix} 
$$

 We now turn to the remaining curvature equations and Bianchi identities. To make life easy, we define the  weighted derivative operators, [4] section 4.14.\begin{mapleinput}
\mapleinline{active}{1d}{thorn:=X->diffsbs(diff(X,r)):}{}
\end{mapleinput}\begin{mapleinput}
\mapleinline{active}{1d}{thorn1:=proc(X,p,q)
 local i;
 D1(X)-p*gamma*X-q*conj(gamma)*X;
 diffsbs(expand(
end proc:}{}
\end{mapleinput}\begin{mapleinput}
\mapleinline{active}{1d}{edth:=proc(X,p,q)
 local i;
 del(X)-p*beta*X-q*conj(alpha)*X;
 diffsbs(expand(
end proc:}{}
\end{mapleinput}\begin{mapleinput}
\mapleinline{active}{1d}{edth1:=proc(X,p,q)
 local i;
 del1(X)-p*alpha*X-q*conj(beta)*X;
 diffsbs(expand(
end proc:}{}
\end{mapleinput} As a check on the calculations, we can run through the curvature equations, [4] (4.12.32), some of which we've used already, some of which will give us $\Psi_3$ and $\Psi_4$. The only ones that do not give zero are
($b'$) and ($c'$), the first of which gives us $\Psi_4$:
\begin{mapleinput}
\mapleinline{active}{1d}{thorn1(sigma1,-3,1)-edth1(kappa1,-3,-1)-sigma1*(rho1+conj(rho1))
+kappa1*(conj(tau)):}{}
\end{mapleinput}\begin{mapleinput}
\mapleinline{active}{1d}{Psi4:=Hlsbs(diffsbs(
\end{mapleinput}
\begin{align*}
{\Psi 4} &:= -8\ L \left(x^a\right)^{3}{z}^{4} Q\left(x^a\right)^{2}
A\left(u\right)^{2}\\
& + \left(\left( -12\,{\frac {r{z}^{7/2}A \left( u \right) }{\sqrt {w}}}-
32\,A( u) ^{2}{z}^{3}w \right) 
Q\left(x^a\right)^{2} -8\,{z}^{2}Q(x^a) \right) L \left(x^a \right)^{2}\\
&+ \Biggl[ \left( -32\,A \left( u \right)^{2}{z}^{2}{w}^{2}-8\, \left( A \left( u \right)  \right) ^{2}{z}^{3}w{\frac {d\,A}{du}} -24\,r{z}^{5/2}\sqrt {w}A \left( u \right)  \right)  
Q \left(x^a\right)^{2}\\ &+ \left( -8\,zw-8{\frac {d\,A}{du}}{z}^{2} \right) Q \left(x^a\right)  \Biggr] L \left(x^a \right)-
8\,{\frac {d\,A}{du}} zw\,Q \left(x^a\right) 
\\ & + \left( -12\,rA \left( u \right) {z}^{3/2}{w}^{3/2}-8\,z{w}^{3} A \left( u \right)^{2}-8A \left( u \right)^{2}{z}^{2}{w}^{2}{\frac {d}{du}}A \left( u \right)  \right)Q(x^a)^{2}
\end{align*}

And ($c'$) gives $\Psi_3$ (as do several others):
\begin{mapleinput}
\mapleinline{active}{1d}{-thorn(kappa1)+conj(tau)*rho1+sigma1*tau:}{}
\end{mapleinput}\begin{mapleinput}
\mapleinline{active}{1d}{Psi3:=diffsbs(
\end{mapleinput}
\begin{align*}
{\Psi 3}&:= \,6\,Q \left(x^a\right) ^{2}L \left(x^a\right)^{2} 
A \left( u \right)^{2}{z}^{{13}/{4}}{w}^{1/4}\\
 & + \left(  \left( 14\, \left( A \left( u \right)  \right) ^{2}{z}^{9/4}{w}^{5/4}+6\,{\frac {A \left( u \right) {z}^{11/4}r}{{w}^{1/4} }} \right)Q \left(x^a\right)^{2}+2\,Q\left(x^a\right) {z}^{5/4}
{w}^{1/4} \right) L \left(x^a\right) \\
 & + \left( 2\,A \left( u\right) ^{2}{z}^{9/4}{w}^{5/4}{\frac {d\,A}{du}} +6\,{z}^{5/4}{w}^{9/4}A \left( u \right)^{2}+6\,A \left( u \right) {z}^{7/4}{w}^{3/4}r \right) Q \left(x^a\right) ^{2}\\ & +2\,{\frac {d\,A}{du}}\,{z}^{5/4}{w}^{1/4}\,Q\left(x^a\right) 
\end{align*}

Now we check the leading terms (in inverse powers of $r$) of our  $\Psi_3$ and $\Psi_4$ 
and compare to [1].\begin{mapleinput}
\mapleinline{active}{1d}{subs(L(u,r,z,w)=Lis,Q(u,r,z,w)=Qis,Hlsbs(Psi4)):}{}
\end{mapleinput}\begin{mapleinput}
\mapleinline{active}{1d}{T4:=subs(r=1/R,
\end{mapleinput}\begin{mapleinput}
\mapleinline{active}{1d}{series(T4,R=0,3) assuming R::positive; }{}
\end{mapleinput}
\[
 -8\,{\frac {d\, A}{du}}zw{R}^{2}+O\left( {R}^{3} \right)
\]
 Here the leading term agrees with [1]. Next $\Psi_3$
\begin{mapleinput}
\mapleinline{active}{1d}{subs(L(u,r,z,w)=Lis,Q(u,r,z,w)=Qis,Hlsbs(Psi3)):}{}
\end{mapleinput}\begin{mapleinput}
\mapleinline{active}{1d}{T3:=subs(r=1/R,
\end{mapleinput}\begin{mapleinput}
\mapleinline{active}{1d}{series(T3,R=0,4) assuming R::positive;}{}
\end{mapleinput}
$$ 2\,\frac{d\,A}{du}{z}^{5/4}{w}^{1/4}{R}^{2}+
8\,{z}^{7/4}{w}^{3/4}A\left( u \right) {R}^{3}+O \left( {R}^{4} \right) $$
 We find that the leading term agrees with [1], but in the second term the powers of 
$z=\zeta$ and $w=\bar\zeta$ are wrong in [1].
We can also check that the Bianchi identities, [4] (4.12.36-39) are satisfied (and they are).

Finally, we turn to the remaining integrability conditions, [3] ($10e$) to ($10h$).
\begin{mapleinput}
\mapleinline{active}{1d}{P110sbs2(diffsbs(Psi2*Hmb-diff(p00,r)-Psi3*Hl(u))); # [3] 10e}{}
\end{mapleinput}
\mapleresult
\begin{maplelatex}
\mapleinline{inert}{2d}{0}{%
\[
0
\]
}

\end{maplelatex}
\begin{mapleinput}
\mapleinline{active}{1d}{Psi3*Hm-Psi2*Hn-2*rho1*p01+2*beta*p00+del(p00): # [3] 10f}{}
\end{mapleinput}\begin{mapleinput}
\mapleinline{active}{1d}{P110sbs2(diffsbs(
\end{mapleinput}
\mapleresult
\begin{maplelatex}
\mapleinline{inert}{2d}{0}{%
\[
0
\]
}

\end{maplelatex}
\begin{mapleinput}
\mapleinline{active}{1d}{Psi3*Hmb-Psi4*Hl(u)+2*sigma1*p01-2*alpha*p00-del1(p00): # [3] 10g}{}
\end{mapleinput}\begin{mapleinput}
\mapleinline{active}{1d}{P110sbs2(diffsbs(
\end{mapleinput}
\mapleresult
\begin{maplelatex}
\mapleinline{inert}{2d}{0}{%
\[
0
\]
}

\end{maplelatex}
\begin{mapleinput}
\mapleinline{active}{1d}{Psi4*Hm-Psi3*Hn-2*kappa1*p01+2*gamma*p00+D1(p00): ## [3] 10h}{}
\end{mapleinput}\begin{mapleinput}
\mapleinline{active}{1d}{Hlsbs(P110sbs2(diffsbs(
\end{mapleinput}
\mapleresult
\begin{maplelatex}
\mapleinline{inert}{2d}{0}{%
\[
0
\]
}

\end{maplelatex}
 So we see that all the homothetic and Killing equations are satisified and we have shown that there is always a Killing vector in these metrics and also always a homothety.

\section{The Calculations (limit cylindrical case)}\label{sc:calclim}

Since neither [1] not [2] give the spin coefficients for the limit cylindrical metric, we will need to calculate them using Maple's {\tt tensor} package.
Note that we use the corrected version of this metric, see [5] equation (26.23)
\begin{mapleinput}
\mapleinline{active}{1d}{with(tensor):}{}
\end{mapleinput}\begin{mapleinput}
\mapleinline{active}{1d}{coord:=[u,r,x,y]:g_c:=array(1..4,1..4,symmetric,sparse):}{}
\end{mapleinput}\begin{mapleinput}
\mapleinline{active}{1d}{g_c[1,1]:=-expand(simplify((b+log(r^2*x^4))/x^2/2)
 assuming r::positive,x::positive);}{}
\end{mapleinput}
\[
{\it g\_c}_{{1,1}}\, := -\,{\frac {b}{2{x}^{2}}}-{\frac {\ln  \left( r \right) }{{x}^{2}}}-2\,{\frac {\ln  \left( x \right) }{{x}^{2}}}
\]
\begin{mapleinput}
\mapleinline{active}{1d}{g_c[1,2]:=1:g_c[3,3]:=-2*r^2:g_c[4,4]:=-2*x^2:g_c[1,3]:=2*r/x:}{}
\end{mapleinput}\begin{mapleinput}
\mapleinline{active}{1d}{g:=create([-1,-1],eval(g_c)):}{}
\end{mapleinput}
Next we calulate all the relevent tensors.
\begin{mapleinput}
\mapleinline{active}{1d}{tensorsGR(coord,g,gup,'detg', 'C1','C2','Rm','Rc', 'R','G','C');}{}
\end{mapleinput}

To calculate the spin coefficients, I use a set of routines available on my web site {\tt http://www.maths.unsw.edu.au/}$\sim${\tt jds/papers.html}
\begin{mapleinput}
\mapleinline{active}{1d}{read `PRcoeff`:}{}
\end{mapleinput}
Now we define the tetrad, with the choice of $m^a$ dictated by the need for the tetrad to be right-handed, so the anti-self duality used in the definition of the
homothetic bivector (see [3]) is satisfied.
\begin{mapleinput}
\mapleinline{active}{1d}{md:=create([-1],vector([0,0,r,-I*x])):mup:=raise(gup,md,1):}{}
\end{mapleinput}
\begin{mapleinput}
\mapleinline{active}{1d}{mbd:=create([-1],vector([0,0,r,I*x])):mbup:=raise(gup,mbd,1):}{}
\end{mapleinput}
\begin{mapleinput}
\mapleinline{active}{1d}{ld:=create([-1],vector([1,0,0,0])):lup:=raise(gup,ld,1):}{}
\end{mapleinput}
\begin{mapleinput}
\mapleinline{active}{1d}{nd:=create([-1],vector([g_c[1,1]/2,1,2*r/x,0])):nup:=raise(gup,nd,1):}{}
\end{mapleinput}
\begin{mapleinput}
\mapleinline{active}{1d}{his:=linalg[stackmatrix](ld[compts],nd[compts],md[compts],mbd[compts]):}{}
\end{mapleinput}
\begin{mapleinput}
\mapleinline{active}{1d}{h:=create([1,-1],op(his)):}{}
\end{mapleinput}
Using the routines {\tt PRspin} and {\tt PRcrv} from the {\tt PRcoeff} file we 
calulate the spin coefficients, and curvature components.
\begin{mapleinput}
\mapleinline{active}{1d}{spins:=PRspin(g,h,C2,coord):}{}
\end{mapleinput}
\begin{mapleinput}
\mapleinline{active}{1d}{crv:=PRcurve(g,h,C,Rc,coord);}{}
\end{mapleinput}
From these two calculations we find that the non-zero spin coefficients are
$$ \tau = \beta=\tau'=-\frac{1}{2rx},\quad
\rho=\sigma=-\frac{1}{2r},\quad \gamma = -\frac{1}{4rx^2},\quad
\rho'=\sigma'=\frac{b+\log(r^2x^4)}{8rx^2};$$
and the non-zero curvature components are
$$ \Psi_1 =\frac{1}{2r^2x} ;\qquad \Psi_2=\frac{1}{2r^2x^2};
\qquad \Psi_3 =  \frac{b+\log(r^2x^4)}{8r^2x^3}.
$$

Now we define the derivative operators $D$, $D'$, $\delta$ and $\delta'$
\begin{mapleinput}
\mapleinline{active}{1d}{De:=XX->add(lup[compts][i]*diff(XX,coord[i]),i=1..4):}{}
\end{mapleinput}\begin{mapleinput}
\mapleinline{active}{1d}{D1:=XX->add(nup[compts][i]*diff(XX,coord[i]),i=1..4):}{}
\end{mapleinput}\begin{mapleinput}
\mapleinline{active}{1d}{del:=XX->add(mup[compts][i]*diff(XX,coord[i]),i=1..4):}{}
\end{mapleinput}
\begin{mapleinput}
\mapleinline{active}{1d}{del1:=XX->add(mbup[compts][i]*diff(XX,coord[i]),i=1..4):}{}
\end{mapleinput} 

 Now to find the Killing vectors. Using [3] ($6a$),($6c$) and (11) gives $\Hl=\Hl(u)$. Then from [3] ($10a$) we get $\phi_{11}$, and find that $\phi_{11}^0(u,x,y)$, the integration constant, is real by [3] ($6g$), 
which also gives $\Hm$. So
\begin{mapleinput}
\mapleinline{active}{1d}{Hm:=-r*p110(u,x,y)+I*Hm0(u,x,y):Hmb:=-r*p110(u,x,y)-I*Hm0(u,x,y):}{}
\end{mapleinput}

\begin{mapleinput}
\mapleinline{active}{1d}{p11:=Hl(u)/2/x/r+p110(u,x,y); # note that p110 is real}{}
\end{mapleinput}
\[
{\it p11}\, := {\frac {{\it Hl} \left( u \right) }{2xr}}+{\it p110} \left( u,x,y \right) 
\]

 We also solve [3] ($10b$) for $\phi_{01}$.
\begin{mapleinput}
\mapleinline{active}{1d}{crv[Psi1]*Hm-2*spins[sigma]*p01-(spins[beta]-spins[alpha1])*p11+del(p11):}{}
\end{mapleinput}\begin{mapleinput}
\mapleinline{active}{1d}{p01:=expand(solve(
\end{mapleinput}
\[
{\it p01}\, := -{\frac {{\it p110}\left(u,x,y\right)}{2x}}
-i{\frac {{\it Hm0} \left( u,x,y \right) }{2xr}}-{\frac {{\it Hl} \left( u \right) }{4{x}^{2}r}}+\frac{1}{2}{\frac {\partial }{\partial x}}{\it p110} \left( u,x,y \right) -i{\frac {r}{2x}}{\frac {\partial }{\partial y}}{\it p110} \left( u,x,y \right)
\]

Now looking at [3] ($8a$), using the fact that $\kappa=0$:
\begin{mapleinput}
\mapleinline{active}{1d}{diff(p01,r)+del1(p11)-2*spins[rho]*p01-
(spins[tau1]+spins[alpha]-spins[beta1])*p11:}{}
\end{mapleinput}\begin{mapleinput}
\mapleinline{active}{1d}{expand(
\end{mapleinput}

\[
-{\frac {3i}{2x}}{\frac {\partial }{\partial y}}{\it p110} \left( u,x,y \right) 
\]

 So $\phi_{11}^0$ is independent of $y$. 
Now looking at [3] ($6b$):
\begin{mapleinput}
\mapleinline{active}{1d}{D1(Hl(u))+2*spins[epsilon1]*Hl(u)+spins[tau]*Hm+spins[tau]*Hmb+p01
+subs(I=-I,p01)-psi:}{}
\end{mapleinput}\begin{mapleinput}
\mapleinline{active}{1d}{collect(
\end{mapleinput}
\[
{\frac {d}{du}}{\it Hl} \left( u \right) +{\frac {\partial }{\partial x}}{\it p110} \left( u,x,y \right) -\psi
\]

 So we solve this for $\phi_{11}^0$, recalling that $\phi_{11}^0$ is independent of $y$, and use it to redefine $\phi_{11}$ , $\phi_{01}$ and $\Hm$.
\begin{mapleinput}
\mapleinline{active}{1d}{p11:=Hl(u)/2/x/r+(psi-diff(Hl(u),u))*x+p0(u): # note that p0 is real}{}
\end{mapleinput}\begin{mapleinput}
\mapleinline{active}{1d}{p01:=expand(subs(p110(u,x,y)=(psi-diff(Hl(u),u))*x+p0(u),p01)):}{}
\end{mapleinput}\begin{mapleinput}
\mapleinline{active}{1d}{Hm:=expand(subs(p110(u,x,y)=(psi-diff(Hl(u),u))*x+p0(u),Hm));}{}
\end{mapleinput}
\mapleresult
\begin{maplelatex}
\mapleinline{inert}{2d}{Hm := -r*x*psi+r*x*diff(Hl(u), u)-r*p0(u)+I*Hm0(u, x, y)}{%
\[
{\it Hm}\, := \,-rx\psi+rx{\frac {d}{du}}{\it Hl} \left( u \right) -r{\it p0} \left( u \right) +i{\it Hm0} \left( u,x,y \right) 
\]
}

\end{maplelatex}
\begin{mapleinput}
\mapleinline{active}{1d}{Hmb:=expand(subs(p110(u,x,y)=(psi-diff(Hl(u),u))*x+p0(u),Hmb)):}{}
\end{mapleinput}Turning to [3] ($10c$)
\begin{mapleinput}
\mapleinline{active}{1d}{crv[Psi2]*Hl(u)-crv[Psi1]*Hmb-2*spins[rho]*p01-(spins[alpha]-spins[beta1])*p11
+del1(p11):}{}
\end{mapleinput}\begin{mapleinput}
\mapleinline{active}{1d}{expand(
\end{mapleinput}
\mapleresult
\begin{maplelatex}
\mapleinline{inert}{2d}{0}{%
\[
0
\]
}

\end{maplelatex}
 Now the right hand side of [3] ($6d$) is 
\begin{mapleinput}
\mapleinline{active}{1d}{-2*spins[epsilon]*Hn-spins[tau1]*Hm-spins[tau1]*Hmb+p01+subs(I=-I,p01)+psi:}{}
\end{mapleinput}\begin{mapleinput}
\mapleinline{active}{1d}{expand(
\end{mapleinput}
\[
{\frac {d}{du}}{\it Hl} \left( u \right) -2\,{\frac {{\it p0} \left( u \right) }{x}}-{\frac {{\it Hl} \left( u \right) }{2{x}^{2}r}}
\]
 This is $D\Hn$, so we integrate 
\begin{mapleinput}
\mapleinline{active}{1d}{int(
\end{mapleinput}
\[
 r{\frac {d}{du}}{\it Hl} \left( u \right)-2\,{\frac {r }{x}}{\it p0} \left( u \right) -{\frac {{\it Hl} \left( u \right) \ln  \left( r \right) }{{2x}^{2}}}
\]
\begin{mapleinput}
\mapleinline{active}{1d}{Hn:=
\end{mapleinput} 
Turning to [3] ($6i$),
\begin{mapleinput}
\mapleinline{active}{1d}{del(Hm)+spins[sigma1]*Hl(u)+spins[sigma]*Hn+(spins[alpha1]+spins[alpha])*Hm:}{}
\end{mapleinput} The coefficients of $r$ are independent, so we collect the terms.
\begin{mapleinput}
\mapleinline{active}{1d}{collect(expand(
\end{mapleinput}
\begin{align*}
-\psi&+\frac{1}{2}{\frac {d}{du}}{\it Hl}\left(u\right) -{\frac {{\it p0} \left( u \right) }{2x}}-{\frac {1}{2x}}{\frac {\partial }{\partial y}}{\it Hm0} \left( u,x,y \right) \\ &
-\left( \frac{1}{2}i{\frac {\partial }{\partial x}}{\it Hm0} \left( u,x,y \right) +{\frac {{\it Hl} \left( u \right) b}{8{x}^{2}}}+
{\frac {{\it Hl} \left( u \right) \ln  \left( x \right) }{2{x}^{2}}}
-{\frac {i{\it Hm0} \left( u,x,y \right) }{2x}}+\frac{1}{2}{\it Hn0} \left( u,x,y \right)  \right) {r}^{-1}
\end{align*}

 The imaginary part of the $r^{-1}$  term implies $\Hm^0=xf(u,y)$,
for some function $f(u,y)$, so:
\begin{mapleinput}
\mapleinline{active}{1d}{X:=collect(subs(Hm0(u,x,y)=x*f(u,y),
\end{mapleinput}\begin{mapleinput}
\mapleinline{active}{1d}{Y:=solve(op(5,X),Hn0(u,x,y));}{}
\end{mapleinput}
\[
Y\, := -{\it Hl}\left(u\right)\,{\frac{ b+4\,\ln\left(x \right)}{4{x}^{2}}}
\]
\begin{mapleinput}
\mapleinline{active}{1d}{expand(subs(Hn0(u,x,y)=Y,X));}{}
\end{mapleinput}
\[
-\psi+\frac{1}{2}{\frac {d}{du}}{\it Hl} \left( u \right) -
\,{\frac {{\it p0} \left( u \right) }{2x}}-\frac{1}{2}{\frac {\partial }{\partial y}}f \left( u,y \right) 
\]
\begin{mapleinput}
\mapleinline{active}{1d}{XX:=rhs(dsolve(
\end{mapleinput}
\[
{\it XX}\, := \, \left( -2\,y\psi+y{\frac {d}{du}}{\it Hl} \left( u \right) -{\frac {y{\it p0} \left( u \right) }{x}}+{\it \_F1} \left( u \right)  \right) x
\]
Where $\_F1$ is an arbitrary function. This $XX$ is $\Hm^0$. So 
\begin{mapleinput}
\mapleinline{active}{1d}{Hm:=collect(subs(Hm0(u,x,y)=XX,Hm),[r,x,y]);
Hmb:=collect(subs(Hm0(u,x,y)=XX,Hmb),[r,x,y]):}{}
\end{mapleinput}
\[
{\it Hm}\, := \, \left[  \left( -\psi+{\frac {d}{du}}{\it Hl} \left( u \right)  \right) x-{\it p0} \left( u \right)  \right] r+ \left[ i \left( -2\,\psi+{\frac {d}{du}}{\it Hl} \left( u \right)  \right) y+i{\it \_F1} \left( u \right)  \right] x-iy\,{\it p0} \left( u \right) 
\]
\begin{mapleinput}
\mapleinline{active}{1d}{Hn:=subs(Hn0(u,x,y)=Y,Hn);}{}
\end{mapleinput}
\[
{\it Hn}\, := \, \left( {\frac {d}{du}}{\it Hl} \left( u \right)  \right) r-2\,{\frac {{\it p0} \left( u \right) r}{x}}-
\frac {{\it Hl}\left( u \right)}{4{x}^{2}} 
\Bigl[2\ln  \left( r \right) + b+4\,\ln  \left( x \right)  \Bigr]
\]
\begin{mapleinput}
\mapleinline{active}{1d}{p01:=expand(subs(Hm0(u,x,y)=XX,p01));}{}
\end{mapleinput}
\[
{\it p01}\, := {\frac {{\it p0} \left( u \right) }{2x}}+{\frac {iy\psi}{r}}-{\frac {iy}{2r}}{\frac {d}{du}}{\it Hl} \left( u \right) 
+{\frac {iy}{2rx}}{\it p0} \left( u \right) 
-{\frac {i}{2r}}{\it \_F1} \left( u \right) -{\frac {{\it Hl} \left( u \right) }{4{x}^{2}r}}
\]
 Returning to the integrability conditions, we look at [3] (10$d$)
\begin{mapleinput}
\mapleinline{active}{1d}{eq10d:=crv[Psi2]*Hm-crv[Psi1]*Hn-2*spins[tau]*p01-2*spins[gamma]*p11+D1(p11):}{}
\end{mapleinput}\begin{mapleinput}
\mapleinline{active}{1d}{expand(
\end{mapleinput}
\[
\frac{{\it p0}\left( u \right) }{2{x}^{2}r} +
\frac{d}{du}{\it p0}\left(u\right) - 
x\,\frac{d^2}{du^2}{\it Hl} \left( u \right)
\]

So by comparing coefficients we have
\begin{mapleinput}
\mapleinline{active}{1d}{p0(u):=0;Hl:=x->k0*x+k1;}{}
\end{mapleinput}
\mapleresult
\begin{maplelatex}
\mapleinline{inert}{2d}{p0(u) := 0}{%
\[
{\it p0} \left( u \right) \, := \,0
\]
}
\mapleinline{inert}{2d}{Hl := proc (x) options operator, arrow; k0*x+k1 end proc}{%
\[
 {\it Hl} := \,x\mapsto {\it k0} x + {\it k1}
\]
}

\end{maplelatex}
And a quick check shows that {\tt eqn10d} ([3] ($10d$)) is satisfied.
Next, the conjugate of [3] ($6h$) will give us $\phi_{00}$.
\begin{mapleinput}
\mapleinline{active}{1d}{D1(Hmb)+spins[tau]*Hn+(spins[gamma]+spins[epsilon1])*Hmb:}{}
\end{mapleinput}\begin{mapleinput}
\mapleinline{active}{1d}{p00:=collect(expand(
\end{mapleinput}
\begin{align*}
{\it p00}\, &:= \left( -{\frac {b}{4x}}-{\frac {\ln\left(r\right)}{2x}}
-{\frac {\ln  \left( x \right) }{x}} \right) \psi+ 
\left({\frac {\ln  \left( r \right) }{4r{x}^{3}}}+
{\frac {\ln  \left( x \right) }{2r{x}^{3}}}
+{\frac {b}{8r{x}^{3}}} \right) {\it k1}+ \\[2pt]
&\left({\frac {ub}{8r{x}^{3}}}+{\frac {b}{4x}}-\frac{1}{2x}
+{\frac {\ln  \left( r \right) }{2x}}+{\frac {\ln\left(x\right) }{x}}+
{\frac {u\ln  \left( x \right) }{2r{x}^{3}}}+
{\frac {\ln  \left( r \right) u}{4r{x}^{3}}} \right) {\it k0}-
i \left( {\frac {d}{du}}{\it\_F1} \left( u \right)  \right) x
\end{align*}

 We next check some further integrability conditions, [3] ($10 e$) first.
\begin{mapleinput}
\mapleinline{active}{1d}{crv[Psi3]*Hl(u)-crv[Psi2]*Hmb-2*spins[tau1]*p01+2*spins[epsilon]*p00
+diff(p00,r):}{}
\end{mapleinput}\begin{mapleinput}
\mapleinline{active}{1d}{expand(
\end{mapleinput}
\mapleresult
\begin{maplelatex}
\mapleinline{inert}{2d}{0}{%
\[
0
\]
}

\end{maplelatex}
And then [3] ($8d$).
\begin{mapleinput}
\mapleinline{active}{1d}{e8d:=diff(p00,r)+del1(p01)-spins[rho]*p00-2*spins[tau1]*p01-spins[sigma1]*p11:}{}
\end{mapleinput}\begin{mapleinput}
\mapleinline{active}{1d}{expand(
\end{mapleinput}
\[
-\frac{ix}{2r} 
\left( {\frac {d}{du}}{\it \_F1} \left( u \right)  \right)
\]
 So the integability function $\_F1$ is constant:
\begin{mapleinput}
\mapleinline{active}{1d}{_F1(u):=k3;expand(e8d);}{}
\end{mapleinput}
\mapleresult
\begin{maplelatex}
\mapleinline{inert}{2d}{_F1(u) := k3}{%
\[
{\it \_F1} \left( u \right) \, := \,{\it k3}
\]
}
\mapleinline{inert}{2d}{0}{%
\[
0
\]
}

\end{maplelatex}
 Also [3] ($6e$) is\begin{mapleinput}
\mapleinline{active}{1d}{eqn6e:=expand(D1(Hn)+2*spins[gamma]*Hn);}{}
\end{mapleinput}
\[
{\it eqn6e}\, :=-{\frac {{\it k0}}{2{x}^{2}}}
\]
Thus ${\tt k0}=0$, and the coefficients simplify as follows:
\begin{mapleinput}
\mapleinline{active}{1d}{Hl(u);Hn;Hm;}{}
\end{mapleinput}
\begin{align*}
&{\it k1}\\
-{\frac {\ln  \left( r \right) {\it k1}}{2{x}^{2}}}&-
{\frac {{\it k1}\, \left( b+4\,\ln  \left( x \right)  \right) }{4{x}^{2}}}
\\
-rx\psi &+ \left( -2\,iy\psi+i{\it k3} \right) x
\end{align*}
\begin{mapleinput}
\mapleinline{active}{1d}{p00;p01;p11;}{}
\end{mapleinput}
\begin{align*}
 \left(-{\frac {b}{4x}}-{\frac {\ln  \left( r \right) }{2x}}-
{\frac {\ln  \left( x \right) }{x}} \right) \psi &+ 
\left( {\frac {\ln  \left( r \right) }{r{x}^{3}}}+
{\frac {\ln  \left( x \right) }{2r{x}^{3}}}+{\frac {b}{8r{x}^{3}}} \right) {\it k1}\\[2pt]
{\frac {iy\psi}{r}}&-{\frac {i{\it k3}}{2r}}-{\frac {{\it k1}}{4{x}^{2}r}}
\\
{\frac {{\it k1}}{2rx}}&+x\psi
\end{align*}

All the remaining homothetic equations and integrability equations are 
satisfied, and we are left with the general homothetic vector:
\begin{mapleinput}
\mapleinline{active}{1d}{lin_com(Hl(u),nup,Hn,lup,-Hm,mbup,-Hmb,mup);}{}
\end{mapleinput}
\mapleresult
\begin{maplelatex}
\mapleinline{inert}{2d}{TABLE([index_char = [1], compts = vector([k1, 2*r*psi, -x*psi, 2*y*psi-k3])])}{%
\[
{\it TABLE} \left( [{\it index\_char}=[1],{\it compts}={\it vector} \left( [{\it k1},2\,r\psi,-x\psi,2\,y\psi-{\it k3}] \right) ] \right) 
\]
}
\end{maplelatex}
That is,
$$k_1\partial_u + k_3\partial_y + 
\psi\left(2r\partial_r -x\partial_x + 2y\partial_y \right).$$

\section{Acknowledgments}

Maple is a registered trademark of Waterloo Maple Inc.

\section{References}

\begin{enumerate}
\item[{[1]}]  Newman and Tamburino  1962 {\it J.~Math.~Phys.} {\bf 3} 902
\item[{[2]}]  Collinson C D and French D C 1967 
 {\it J.~Math.~Phys.} {\bf 8} 701
\item[{[3]}] Steele J D 2002 {\it Class. Quantum Grav.} {\bf 19} 259
\item[{[4]}] Penrose R and Rindler W 1984 {\it Spinors and Space-Time} vol 1
(Cambridge: Cambridge University Press)
\item[{[5]}] Stephani H, Kramer D, MacCallum M, Hoensaerlars C and Herlt E 2003
{\it Exact Solutions to Einstein's Field Equations 2nd Edition} (Cambridge: Cambridge University Press)
\item[{[6]}] Fayos F and Sopuerta C F 2001 {\it Class. Quantum Grav.} {\bf 18} 353 
\end{enumerate} 
\end{document}